\documentclass[12pt]{article}  
\usepackage{graphicx}
\usepackage[centertags]{amsmath}
\usepackage[english]{babel}
\usepackage{amsthm}
\usepackage{mathtools}
\usepackage{fancyhdr}  
\usepackage{setspace}   
\usepackage{caption} 

\usepackage{amsfonts}
\usepackage{amssymb}
\usepackage{multirow}
\usepackage{algorithm}
\usepackage[noend]{algpseudocode}

\title{ \Large \textbf{STAD Research Report 01/2015} \\ \vspace{10mm} Boosted-Oriented Probabilistic Smoothing-Spline Clustering of Series. \\   }

\author{Carmela Iorio*, Gianluca Frasso**, \\
Antonio D'Ambrosio* and Roberta Siciliano***\\
\\\small *Department of Economics and Statistics. 
\\\small University of Naples Federico II\\ 
\small carmela.iorio@unina.it, antdambr@unina.it\\
\\\small **Institut des sciences humaines et sociales,\\
\small M\'ethodes quantitatives en sciences sociales\\
\small Universit\'e de Li\'ege, Belgium\\
\small Gianluca.Frasso@ulg.ac.be\\
\\\small ***Department of Industrial Engineering
\\\small University of Naples Federico II\\
\small roberta@unina.it\\}

\date{March 23, 2015}

\begin{document}
\maketitle
\newpage


\noindent \textbf{Abstract}.
Fuzzy clustering methods allow the objects to belong to several clusters simultaneously, with different degrees of membership. However, a factor that influences the performance of fuzzy algorithms is the value of fuzzifier parameter. In this paper, we propose a fuzzy clustering procedure for data (time) series that does not depend on the definition of a fuzzifier parameter.
It comes from two approaches, theoretically motivated for unsupervised and supervised classification cases, respectively. The first is the Probabilistic Distance (PD) clustering procedure. The second is the well known Boosting philosophy. Our idea is to adopt a boosting prospective for unsupervised learning problems, in particular we face with non hierarchical clustering problems. The aim is to assign each instance (i.e. a series) of a data set to a cluster. We assume the representative instance of a given cluster (i.e. the cluster center)  as a target instance, a loss function as a synthetic index of the global performance and the probability of each instance to belong to a given cluster as the individual contribution of a given instance to the overall solution. The global performance of the proposed method is investigated by various experiments.
\\

\noindent
{\bf Keywords}: Fuzzy Clustering, Boosting, PD clustering

\section{Introduction}
\label{sec5.1}
We propose a fuzzy approach for clustering data (time) series. The goal of clustering is to discover groups so that objects within a cluster have high similarity among them, and at the same time they are dissimilar to objects in other clusters. Many clustering algorithms for time series have been introduced in the literature. Since clusters can formally be seen as subsets of the data set, one possible classification of clustering methods can be according to whether the subsets are fuzzy (soft) or crisp (hard). 
Let $\mathcal{D}$ be a data set consisting of $N$ series ${\{y_1, y_2,..., y_N\}} \subset \mathbb{R}^n$ and let $K$ be an integer, with $2 \leq K < N $, the goal is to partition $\mathcal{D}$ into ${\mathcal{C}_K}$ groups.
Crisp clustering methods are based on classical set theory, and restrict that each object of data set belongs to exactly one cluster.
It means partitioning the data $\mathcal{D}$  into a specified number of mutually exclusive clusters $\mathcal{C}_1, \mathcal{C}_2, ...\mathcal{C}_K$.\\ 
\noindent A hard partition of $\mathcal{D}$ can be defined as a  family of subsets ${\mathcal{C}_k}$ that satisfies the following properties (Bezdek, 1981):
\[
\bigcup_{k=1}^{K} \mathcal{C}_k= \mathcal{D},
\]
\[
\mathcal{C}_k \cap \mathcal{C}_h =\emptyset, 	\textrm{      }{k \neq h}
\]
\[ 
\emptyset \subset \mathcal{C}_k \subset \mathcal{D},	\textrm{       }{1 \leq k \leq K.} 
\]
Let $\mu_{ik}$ be the membership function and let $\mathbf{U}=[\mu_{ik}]$ be the $N \times K$ partition matrix. The elements of $\mathbf{U}$ must satisfy the following conditions:
\[
\mu_{ik} \in {\{0,1\}}, \textrm{     } 1 \leq k \leq K, \textrm{     } 1 \leq i \leq N;
\]
\[
\sum_{k=1}^{K}\mu_{ik}=1;
\]
\[
0 < \sum_{i=1}^{N} \mu_{ik} < N.
\]
The $k^{th}$ column of $\mathbf{U}$ contains value of $\mu_{ik}$ of the $k^{th}$ subset $\mathcal{C}_k$ of $\mathcal{D}$.\\
In a hard partition, $\mu_k(y_i)$ is the indicator function:

\[\mu_k(y_i)= \left\{
  \begin{array}{lr}
    1, & \mbox{if } y_i \in \mathcal{C}_k ;\\
    0, & \mbox{otherwise}
  \end{array}
\right.
\]

\noindent Following Bezdek (1981) the hard partionining space is thus defined by:
\[
M_c=\{{\mathbf{U} \in \mathbb{R}^{K \times n}}|{\mu_{ik} \in {\{0,1\}}}, \forall i,k;{\sum_{k=1}^{K}\mu_{ik}=1 ,\forall i ,0 < \sum_{i=1}^{N} \mu_{ik} <N, \forall k}\}, 
\]
$M_c$ being the space of all possible hard partition matrices for $\mathcal{D}$.\\
Generalizing the crisp partition, $\mathbf{U}$ is a fuzzy partitions of $\mathcal{D}$ with elements $\mu_{ik}$ of the partition matrix bearing real values in $[0,1]$ (Kaufman and Rousseeuw, 2009).

\noindent The idea of fuzzy set was conceveid by Zadeh (2009). Fuzzy clustering methods do not assign objects to a cluster but suggest degrees of membership to each group. The larger is the value of the membership value for a given object with respect to a cluster, the larger is the probability of that object to be assigned to that cluster.\\
\noindent Similarly to crisping conditions, Ruspini (1970) defined the following fuzzy partition properties:
\[
\mu_{ik} \in [0,1], \textrm{       } 1 \leq k \leq K, \textrm{      } 1 \leq i \leq N;
\]
\[
\sum_{k=1}^{K}\mu_{ik}=1;
\]
\[
 0 < \sum_{i=1}^{N} \mu_{ik} < N. 
\]
\noindent The fuzzy partitioning space is the set:
\[
M_f=\{{\mathbf{U} \in \mathbb{R}^{K \times n}}|{\mu_{ik} \in [0,1]}
, \forall i,k;{\sum_{k=1}^{K}\mu_{ik}=1 ,\forall i ,0 < \sum_{i=1}^{N} \mu_{ik} < N, \forall k}.\} 
\]

\noindent Several clustering criteria have been proposed to identify fuzzy partition in $\mathcal{D}$. Among these proposals, the most popular method is  fuzzy $c$-means.
 
\noindent Proposed by Dunn (1973) and developed by Bezdek (1981), fuzzy $c$-means considers each data point as a possible member of multiple clusters with a membership value.
This algorithm is based on minimization of the following objective function:

\begin{equation}\label{eq5.7}
J_m=\sum_{i=1}^{N} \sum_{k=1}^{K} (\mu_{ik})^m\lVert y_i-c_k\rVert^2\\
\end{equation}

\noindent s.t. \\

\[
  \begin{array}{lr}
	\vspace{0.3cm}
    \mu_{ik} \in [0,1], \textrm{  } \forall  i,k;\\ \vspace{0.3cm}
    \sum_{k=1}^{K} \mu_{ik}=1; \\ \vspace{0.3cm}
		 0 < \sum_{i=1}^{N} \mu_{ik} < N.
  \end{array}
\]

\noindent In the equation \eqref{eq5.7}, $m$ is any real number greater than 1, $\mu_{ik}$ is the degree of membership of $y_i$ in the cluster $k$ and $\lVert \cdot \rVert$  is any norm expressing the similarity between any measured data and the center. The parameter $m$ is called \textit{fuzzifier} or \textit{weighting coefficient}.
To perform fuzzy partitioning, the number of clusters and the weighting coefficient have to be choosen. The procedure is carried out through an iterative optimization of the objective function shown above, with the update of membership value $\mu_{ik}$ and the cluster centers $c_k$ by solving:
\begin{eqnarray} \label{centr}
c_k=\frac{\sum_{i=1}^{N} (\mu_{ik})^m y_i } {\sum_{i=1}^{N} (\mu_{ik})^m},&  k=1,\ldots, K.
\end{eqnarray}

\begin{eqnarray} \label{uik}
\mu_{ik}=\left(  \sum_{h=1}^{K}\left(\frac{\sqrt[(m-1)]{\lVert y_i - c_k \rVert^2}}{\sqrt[(m-1)]{\lVert y_i - c_h \rVert^2}}\right) \right) ^{-1} & i= 1,\ldots, N; & k= 1,\ldots,K.
\end{eqnarray}

The loop will stop when
\[
 max_{ik} \lvert \mu_{ik}^{(l+1)} - \mu_{ik}^{(l)} \rvert < \varepsilon,
\]

\noindent where $\varepsilon$ is a small number for stopping the iterative procedure, and $l$ indicates the iteration steps. The algorithm is synthesized in box \ref{algfcm}.

\begin{algorithm}[h]
 \floatname{algorithm}{Box}
\caption{Fuzzy\textit{c}-means algorithm}
\label{algfcm}
\begin{algorithmic}
\State Initialize: $K$ = number of centers, $m, (1 < m < \infty)$, $\varepsilon$ = a small threshold. Set the counter $l = 1$ and initialize the matrix of the fuzzy $c-$ partitions $U= [u_{ik}^{(l)}]$.
\While {$\lVert U^{(l+1)} - U^{(l)} \rVert > \varepsilon$}
\State - Calculate the cluster center, $c_k^{ (l)} $ by using equation \eqref{centr}. 
\State - Update the membership matrix $U=[\mu_{ik}]$ by using equation \eqref{uik}, if $y_i \neq c_k ^{(l)}$, 
\State ${\textrm{  }}$ otherwise set $u_{ik}=1$ if $l=i$ or set  $u_{ik}=0$ if $l\neq i$.
\State - Compute $\lvert U^{(l+1)} - U^{(l)}\rvert$.
\If  {$\lvert U^{(l+1)} - U^{(l)} \rvert > \varepsilon$ }
\State - Set $l=l+1$
\EndIf
\State \textbf{end if}
\EndWhile 
\State \textbf{end while}
\State \textbf{output}: estimated centers $\widehat{c}_k$, membership matrix $U$.  
\end{algorithmic}
\end{algorithm}


\noindent One of limitations of fuzzy $c$-means clustering is the value of fuzzifier $m$.
A large fuzzifier value tends to mask outliers in data sets, i.e. the larger $m$, the more clusters share their objects and viceversa. For $m\rightarrow\infty$ all data objects have identical membership to each cluster, for $m=1$, the method becomes equivalent to $k$-means. The role of the weighting exponent has been well investigated in literature.

\noindent Pal and Bezdek (1995) suggested taking $m \in [1.5, 2.5]$.

\noindent Demb\'el\'e and Kastner (2003) obtain the fuzzifier with an empirical method calculating the coefficient of variation of a function of the distances between all objects of the entire datset.

\noindent Yu \textit{et al.} (2004) proposed a theoretical upper bound for $m$ that can prevent the sample mean from being the unique optimizer of a fuzzy $c$-means objective functions.

\noindent Futschik and Carlisle (2005) search for a minimal fuzzifier value for which the cluster anlysis of the randomized data set produces no meaningful results, by comparing a modified partitions coefficient for different values of both parameters.  

\noindent Schw\"ammle and Jensen (2010) showed that the optimal fuzzfier takes values far from the its frequently used value equal to $2$. The authors introduced a method to determine the value of the fuzzifier without using the current working data set. Then for high dimensional ones, the fuzzifier value depends directly on the dimension of data set and its number of objects. For low dimensional data set with small number of objects, the authors reduce  the search space to find the optimal value of the fuzzifier. According to the authors, this improvement helps choosing the right parameter and saving computational time when processing large data set.

\noindent On the basis of a robust selection analysis of the algorithm, Wu (2012) founds that a large value of $m$ will make fuzzy $c$-means algorithm more robust to noise and outliers. The author suggested to use value of fuzzifier ranging between 1.5 and 4. \\
\noindent Since the weighting coefficient determines the fuzziness of the resulting  classification, we propose a method that is independent from the choice of the fuzzifier. It comes from two approaches, theoretically motivated for unsupervised and supervised classification cases respectively. The first is the Probabilistic Distance (PD) clustering procedure defined by Ben Israel and Iyigun (2008). The second is the well known Boosting philosophy. From the PD approach we took the idea of determining the probabilities of each series to any of the $k$ clusters. As this probability is unequivocally related to the distance of each series from the centers, there are no degrees of freedom in determine the membership matrix. From the Boosting approach (Freund and Schapire, 1997) we took the idea of weighting each series according some measure of badness of fit in order to define an unsupervised learning process based on a weighted re-sampling procedure. As a learner for the boosting procedure we use a smoothing spline approach. Among the smoothing spline techniques, we chose the penalized spline approach (Eilers and Marx, 1996) because of its flexibility and computational efficiency. This paper is organized as follows: Section \ref{sec5.2} contains our proposal, in Section \ref{sec5.3} the results of some experimental evaluation studies are carried out and some concluding remarks are presented in Section \ref {sec5.4}.

\section{Boosted-oriented probabilistic clustering of time series}
\label{sec5.2}
%
%
%
\subsection{The key idea}
The boosting approach is based on the idea that a supervised learning algorithm (weak learner) improves its performance by learning from its errors (Freund and Schapire, 1997). It consists of an ensemble method that work with a resampling procedure (Dietterich, 2000). The general idea is to run several times the supervised learning algorithm and assigning a weight to each instance of a data set that governs the resampling (with replacement) process during the iterations. The weights are set in such a way that the misclassified instances gets a weight larger than the weight assigned to well classified instances. In this way, the probability to be included in the sample during the iterations is higher for those instances for which the supervised learning algorithm returns a wrong classification. There exist boosting algorithms for both classification and regression problems (Freund and Schapire, 1997; Dietterich, 2000; Eibl and Pfeiffer, 2002; Gey and Poggi, 2006). In both cases the weighting system combines a synthetic index of the performance of the supervised learning algorithm with some index that represents the individual contribution of a given instance to the overall solution. Our idea is to adapt the boosting philosophy to unsupervised learning problems, specially to non hierarchical cluster analysis. In such a case there not exists a target variable, but as the goal is to assign each instance (i.e. a series) to a cluster, we have a target instance. In other words, we switch from a target variable to a target instance point of view. We take each cluster center as a representative instance for each series and we assume as a synthetic index of the global performance a loss function to be minimized. The probability of each instance to belong to a given cluster is assumed to be the individual contribution of a given instance to the overall solution. In contrast to the boosting approach, the larger the probability of a given series to be member of a given cluster, the larger the weight of that series in the resampling process. As a learner either a smoothing spline techniques or a regression model can be used. We decided to use a penalized spline smoother because of its flexibility and computational efficiency. To define the probabilities of each series to belong to a given cluster we use the PD clustering approach (Ben Israel and Iyigun, 2008). This approach allows us to define a suitable loss function and, at the same time, to propose a fuzzy clustering procedure that does not depend on the definition of a fuzzifier parameter.

\subsection{P-splines in a nutshell}

P-splines have been introduced by Eilers and Marx (1996) as flexible smoothing procedures combining B-splines (de Boor, 1978) and difference penalties. Suppose to observe a set of data $\{x, y \}_{j = 1}^{n}$, where the vector ${ x}$ indicates the independent variable (e.g. time) and ${ y}$ the dependent one. We want to describe the available measurements through an appropriate smooth function. Denote ${ B}_{j}(x;p)$ the value of the $i-th$ B-spline of degree $p$ defined on a domain spanned by equidistant knots (in case of not equally spaced knots our reasoning can be generalized using divided differences). A curve that fits the data is given by $\widehat{y}(x) = \sum_{j=1}^{n}a_{j}{ B}_{j}(x;p)$ where $a_{j}$ (with $j=1,...,n$) are the estimated B-splines coefficients. Unfortunately the curve obtained by minimizing $\|{ y} - { B a}\|^{2}$ w.r.t. ${ a}$ shows more variation than is justified by the data if a dense set of spline functions is used. To avoid this overfitting tendency it is possible to estimate ${ a}$ using a generous number of bases in a penalized regression framework
\begin{equation}\label{eq1_3}
\widehat{{ a}} = \mathop{\mbox{argmin}}_{{ a}} \|{ y} -  { Ba} \|^{2} + \lambda\|{Da} \|^{2},
\end{equation}
where ${D}$ is a $d^{th}$ order difference penalty matrix and $\lambda$ is a smoothing parameter. Second or third order difference penalties are suitable in many applications.\\ 
The optimal spline coefficients follow from \eqref{eq1_3} as:
\begin{equation}\label{eq1_4}
\widehat{{ a}} = ({ B}^{\top} { B} + \lambda { D}^{\top} { D})^{-1} { B}^{\top} { y}.
\end{equation}

\noindent The smoothing parameter $\lambda$ controls the trade-off between smoothness and goodness of fit. For $\lambda \rightarrow \infty$ the final estimates tend to be constant while for $\lambda \rightarrow 0$ the smoother tends to interpolate the observations.\\
Popular methods for smoothing parameter selection are the Akaike Information Criterion and Cross Validation. AIC estimates the predictive log likelihood, by correcting the log likelihood of the fitted model ($\Lambda$) by its effective dimension ($\mbox{ED}$): $\mbox{AIC} = 2\mbox{ED} - 2\Lambda$. Following Hastie and Tibshirani (1990) we can compute the effective dimension as $\mbox{ED} = \mbox{tr}[({ B}^{\top} { B} + \lambda { D}^{\top}{ D})^{-1}{ B}^{\top}{ B}]$ for the P-spline smoother and
\[
2\ell = -2n\ln\hat{\sigma}^{2}\sum_{j=1}^{n}\frac{(y_{j}-\hat{y}_{j})^{2}}{\hat{\sigma}_{0}^{2}},
\]
\noindent where $\hat{\sigma}$ is the maximum likelihood estimate of $\sigma$. But $\hat{\sigma}^{2}=\sum_{i}(y_{j}-\hat{y}_{j}^{2})^{2}/n$, so the second term of $\ell$ is a constant. Hence the AIC can be written as
\[
\mbox{AIC}(\lambda) = 2\mbox{ED} + 2n\ln\hat{\sigma}.
\]
\noindent The optimal parameter is the one that minimizes the value of $AIC(\lambda)$.

\noindent LOO-CV chooses the value of $\lambda$ that minimizes
\[
\mbox{CV}(\lambda) = \sum_{i=j}^{n} \left[\frac{y_{j} - \widehat{y}_{j}}{1 - h_{jj}} \right]^{2},
\]
\noindent where $h_{jj}$ is the $j$th diagonal entry of ${ H} = { B} ({ B}^{\top}{ B} + \lambda { D}^{\top}{ D})^{-1}{ B}^{\top}$.

\noindent Analogous to CV is the generalized cross validation measure (Whaba, 1990)
\[
\mbox{GCV}(\lambda) = \sum_{j=1}^{n} \left[\frac{y_{j} - \widehat{y}_{j}}{n - \mbox{ED}} \right]^{2},
\]
\noindent where $\mbox{ED} = \mbox{tr}({ H})$.
\noindent In analogy with cross validation we select the smoothing parameter that minimizes
$\mbox{GCV}(\lambda)$.

\noindent All these selection procedures suffer of two drawbacks: 1) they require the computation of the effective model dimension which can become time consuming for long data series, and 2) they are sensitive to serial correlation in the noise around the trend. The L-curve (Hansen, 1992) and the derived V-curve criteria (Frasso and Eilers, 2015) overcome these hitches. The L-curve is a parameterized curve comparing the two ingredients of every regularization or smoothing procedure: badness of the fit and roughness of the final estimate. For a P-spline smoother, the following quantities can be defined
  \begin{equation*}
  \{\omega(\lambda); \theta(\lambda) \} = \{ \|\mathbf{y} - \mathbf{B} \widehat{\mathbf{a}}\|^{2}; \| \mathbf{D}  \widehat{\mathbf{ a}} \|^{2}\}.
  \end{equation*}
The L-curve is obtained by plotting $\psi(\lambda) = \log(\omega)$ against $\phi(\lambda) = \log(\theta)$. This plot typically shows a L-shaped curve and the optimal amount of smoothing is located in the corner of the ``L'' by maximizing the local curvature measure.
The V-curve criterion offers a valuable simplification of the searching criterion by requiring the minimization of the Euclidean distance between the adjacent points on the L-curve and, like in plots of AIC or GCV, the graphical presentation of the V-curve has an axis for $\lambda$ that can be read off. The V-curve criterion is computed as follows:
\begin{equation}\label{VC}
\mbox{V}(\lambda) = \sqrt{\left(\frac{d \psi}{d \lambda}\right)^2 + \left(\frac{d \phi}{d \lambda}\right)^2}.
\end{equation} 

\subsection{PD clustering approach}

Let $\mathcal{D}$ be a dataset consisting of $N$ series ${\{y_1, y_2,..., y_N\}} \subset \mathbb{R}^n$ and let  $\mathcal{C}_k$ be $k^{th}$ cluster, with $k \in (1,K)$,  partitioning  $\mathcal{D}$. We suppose that each series has the same domain of length $n$.\\
\noindent At each cluster $\mathcal{C}_k$ is associated a cluster center $c_k$, with $k=1, ..., K$.\\ 
Let $d_{i,k}=d(y_i,c_k)$ be a distance function of the i$^{th}$ series from the k$^{th}$ cluster center.\\
Let $P_{i,k}=P(y_i,\mathcal{C}_k)$ be the probability of the i$^{th}$ series belonging to the k$^{th}$ cluster. \\
For each series $y \in \mathcal{D}$ and each cluster  $\mathcal{C}_k$, we assume the following relation between probabilities and distances (Ben Israel and Iyigun, 2008):
\begin{eqnarray}\label{principlepd}
P_{i,k} d_{i,k}= constant.
\end{eqnarray}

\noindent The constant in \eqref{principlepd} only depends on series $y$ and it is independent of the cluster $k$. Equation \eqref{principlepd} allows to to define the membership probabilities as (Heiser, 2004; Ben-Israel and Iyigun, 2008)

\begin{eqnarray} \label{probs}
P_{i,k}=\frac{\prod_{i \neq j} d_{j,k}}{\sum_{k=1}^{K}\prod_{i\neq j}d_{i,k}}.
\end{eqnarray}

\subsection{The algorithm}

Since the probabilities as defined in equation \eqref{probs} sum up to one among the clusters, we use the quantity $\prod_{k=1}^K P_{i,k}$ as a measure of compliance representation of the $i-th$ series with respect to the overall solution of the clustering procedure. It is easy to note that $\prod_{k=1}^K P_{i,k}=0$ if the $i-th$ series coincides with the $k-th$ cluster center, as well as $\prod_{k=1}^K P_{i,k}=K^{-1}$ if there is maximum uncertainty in assigning the $i-th$ series to any cluster center. For this reason we use as measure of cluster compliance solution the quantity

\begin{eqnarray}\label{solpd}
BC=\frac{1}{N}\sum_{i=1}^N (\prod_{k=1}^K P_{i,k}) K^K.
\end{eqnarray}

\noindent Equation \eqref{solpd} is a synthetic uncertainty clustering measure: the lower its value, the better the solution. It equals zero when there is a perfect solution (i.e., each series has probability equal to one to belong to some cluster center). The maximum possible value of equation \eqref{solpd} is $1$, when each series has probability equal to $K^{-1}$ to belong to each of the $K$ cluster. The $BC$ index allows to compare the overall clustering solution when the number $K$ of the clusters differs. \\

\noindent From equation \ref{solpd} we define the following loss function to be minimized as
\begin{equation} \label{loss}
\beta=\sum_{i=1}^N (\prod_{k=1}^K P_{i,k}) K^K .
\end{equation}

\noindent Let $\gamma_{i,k}={d_{i,k}}/{max_{k=1}^K d_{i,k}}$ be the contribution of the $i-th$ series to generate the $k-th$ cluster.\\
\noindent Let $\mathbf{\Gamma}$ be a $N \times K$ indicator matrix whose entries are $1$ if $P_{i,k} > P_{i,h}$ ($k,h=1,\ldots,K$, $k\neq h$) and $-1$ otherwise.\\
We define the weight of the $i-th$ series for the $k-th$ cluster as
\[
w_{i,k}=\beta^{\gamma_{i,k}\Gamma_{i,k}}.
\]

\noindent For each cluster $k$, the weights are first normalized in this way:
\[
w_{i,k}^{\bullet}=\frac{w_{i,k}}{\sum_{h=1}^K w_{i,h}},
\]

\noindent then within each cluster we set

\begin{equation} \label{weights}
W_{i,k}=\frac{w_{i,k}^{\bullet}}{\sum_{i=1}^N w_{i,k}^{\bullet}}.
\end{equation}

\noindent For each cluster $k$, a sample $\mathcal{L}^{(k)}$ is extracted with replacement from $\mathcal{D}$, taking in account equation \eqref{weights}. Then the cluster centers $\widehat{c}_k =  { B} \widehat{a}$, $k=1,\ldots,K$ are estimated by using a P-spline smoother. These centers are then used to compute the membership probabilities according to equation \eqref{probs} for the next iteration. The cluster centers are re-estimated and adaptively updated with an optimal spline smoother.\\
The choice of the metric depends on the nature of the series, the optimal P-spline smoothing procedure frames our approach in the class of model-based clustering techniques but any suitable smoother can be adopted. Box \ref{alg} shows the pseudo-code of our the Boosted-Oriented Smoothing Spline Probabilistic Clustering algorithm.

\begin{algorithm}[t!]
 \floatname{algorithm}{Box}
\caption{Boosted-oriented smoothing-spline probabilistic clustering of time series}
\label{alg}
\begin{algorithmic}
\State \textbf{input} $\mathcal{D}$
\State \textbf{initialize:} maxiter = maximum number of iterations; $K =$ the number of clusters; $d$ = a suitable distance measure; $c_k$, $k=1:\ldots,K$ random cluster centers.
\For{iter=1:maxiter}
\State - compute the $N \times K$ distance matrix $D=[d_{i,k}]$ $\forall{i,k}$;
\State - compute the membership probabilities $P=[P_{i,k}]$  $\forall{i,k}$ as in equation 
\State $\textrm{    }$\eqref{probs};
\State - compute $\beta^{[iter]}$ as in equation \eqref{loss};
\State - assign the weights to each series for each cluster and compute the 
\State $\textrm{    }$$N \times K$  matrix $W$ as in equation \eqref{weights}; 
\For{$k=1:K$}
\State - extract the sample $\mathcal{L}^{k}$ from $\mathcal{D}$
\State - compute center $\widehat{c}_k^{[iter]} =  { B} \widehat{a}_k$
\EndFor
\State \textbf{end for}
\If{iter $= 1$}
\State - $\widehat{c}_k^{*} =  { B} \widehat{a}_k$
\Else
\For{$k=1:K$}
\State - update cluster centers $\widehat{c}_k^{*} =  { B} \widehat{a}_k^{*}$, 
\State $\textrm{        }$ with $\widehat{a}_k^{*} = ({ B}^{\top} { B} + \lambda { D}^{\top} { D})^{-1} { B}^{\top} { \widehat{c}}_k^{[1:iter]}$
\EndFor
\State \textbf{end for}
\EndIf
\State \textbf{end if}
\EndFor
\State{\textbf{end for}}
\State \textbf{output}: estimated cluster centers $\widehat{c}_k^{*}$, membership probabilities matrix $P$. 

\end{algorithmic}
\end{algorithm}
\noindent The procedure described in box \ref{alg} is repeated a certain number of time due to the sensitivity of final solution to the random choice of cluster center.

\section{Experimental evaluation}
\label{sec5.3} 
\noindent To evaluate  the performance of the proposed algorithm, we conducted three experiments. In estimating the optimal P-splines smoother, always we used the V-curve criterion as in equation \eqref{VC} to select the optimal $\lambda$ parameter, and we used a number of interior knots equal to $\min(\frac{n}{4}; 40)$, in which $n$ is the length of time domain, as suggested by Ruppert (2002). Moreover we need a measure of goodness of fuzzy partitions. To reach this aim, we decided to use a fuzzy variant of the Rand Index proposed by Hullermeier \textit{et al.} (2012).  This index is defined by the complement to $1$ of the normalized sum of degree of discordance. The Rand index developed by Rand (1971) is a \textit{external evaluation measure} to compare the clustering partitions on a set of data. The problem of evaluating the solution of a fuzzy clustering algorithm with the Rand index is that it requires converting the soft partition into a hard one, losing information.

\noindent As shown in Campello (2007), different fuzzy partitions describing different structures in the data may lead to the same crisp partition and then in the same Rand index value. For this reason the Rand index is not appropriate for fuzzy clustering assessment.\\ 
\noindent To overcome this problem H\"ullermeier \textit{et al.} (2012) proposed a generalization of the Rand index for fuzzy partitions.  We recall some essential background.\\
Let $\mathcal{P}=\{\mathcal{P}_1, \ldots, \mathcal{P}_K\} $ be a fuzzy partition of the data set $\mathcal{D}$, each element $y_i \in \mathcal{D}$ is characterized by its membership vector:
\begin{eqnarray}\label{memb_vector}
\mathcal{P}_i = (\mathcal{P}_1(y_i), \mathcal{P}_2(y_i), \ldots, \mathcal{P}_k(y_i), \ldots, \mathcal{P}_K(y_i)) & \in [0,1]^K
\end{eqnarray}
where $\mathcal{P}_k(y_i)$ is the degree membership of the $i-th$ series to the $k-th$ cluster $\mathcal{P}_k$. Given any pair $(y_i, y_i^{'}) \in \mathcal{D}$, Hellermeier \textit{et al.} (2012) defined a fuzzy equivalence relation on $\mathcal{D}$ in terms of similarity measure on the associated membership vectors \eqref{memb_vector}. Generally, this relation is of the form:
\[
E_\mathcal{P}=1-\lVert \mathcal{P}_i- \mathcal{P}_{i'}\rVert
\]
where $\lVert \cdot  \rVert$ represents the $L_1$norm divided by $2$ that constitutes a proper metric on $[0, 1]^K$ and yields value on $[0,1]$. $E_\mathcal{P}$ is equal to $1$ if and only if $y_i$ and $y_i^{'}$ have the same membership pattern and is equal to $0$ otherwise. The basic idea of the authors to reach the fuzzy extension of the Rand index was to generalize the concept of concordance in the following way. \\
Given 2 fuzzy partition, $\mathcal{P}$ and $\mathcal{Q}$ and considering a pair $(y_i, y_i^{'})$ as being concordant as $\mathcal{P}$ and $\mathcal{Q}$ agree on its degree of equivalence, they defined the degree of concordance as

\[
conc(y_i, y_i^{'})=1- \parallel E_\mathcal{P}(y_i, y_i^{'}) - E_\mathcal{Q}(y_i, y_i{'}) \parallel \textrm{  } \in [0,1],
\]
\noindent and degree of discordance as:
\[
disc(y_i, y_i^{'})= \parallel E_\mathcal{P}(y_i, y_i^{'}) - E_\mathcal{Q}(y_i, y_i^{'})\parallel \textrm{  } \in [0,1].
\]

\noindent Finally, the distance measure proposed by H\"ullermeier \textit{et al.} (2012) is defined as the normalized sum of degrees of discordance:
\[
d(\mathcal{P},\mathcal{Q}) = \frac{\sum_{(y_i, y_i^{'})\in \mathcal{D}}\lVert E_\mathcal{P}(y_i, y_i^{'}) - E_\mathcal{Q}(y_i, y_i^{'})\rVert }{N(N-1)/2}
\]
\noindent The direct generalization of the Rand index corresponds to the normalized degree of concordance and it is equal to:
\[
R_E(\mathcal{P}, \mathcal{Q})=1-d(\mathcal{P},\mathcal{Q})
\]
and it reduces to the original Rand index when partitions $\mathcal{P}$ and $\mathcal{Q}$ are non-fuzzy.\\

\noindent As true fuzzy partition, we always computed the true cluster centers with an optimal P-spline smoother, and then we computed the true probabilities by applying equation \eqref{probs}.

\subsection{Simulated data}
\begin{figure}[h]
\centering
\includegraphics[width=1.00\textwidth]{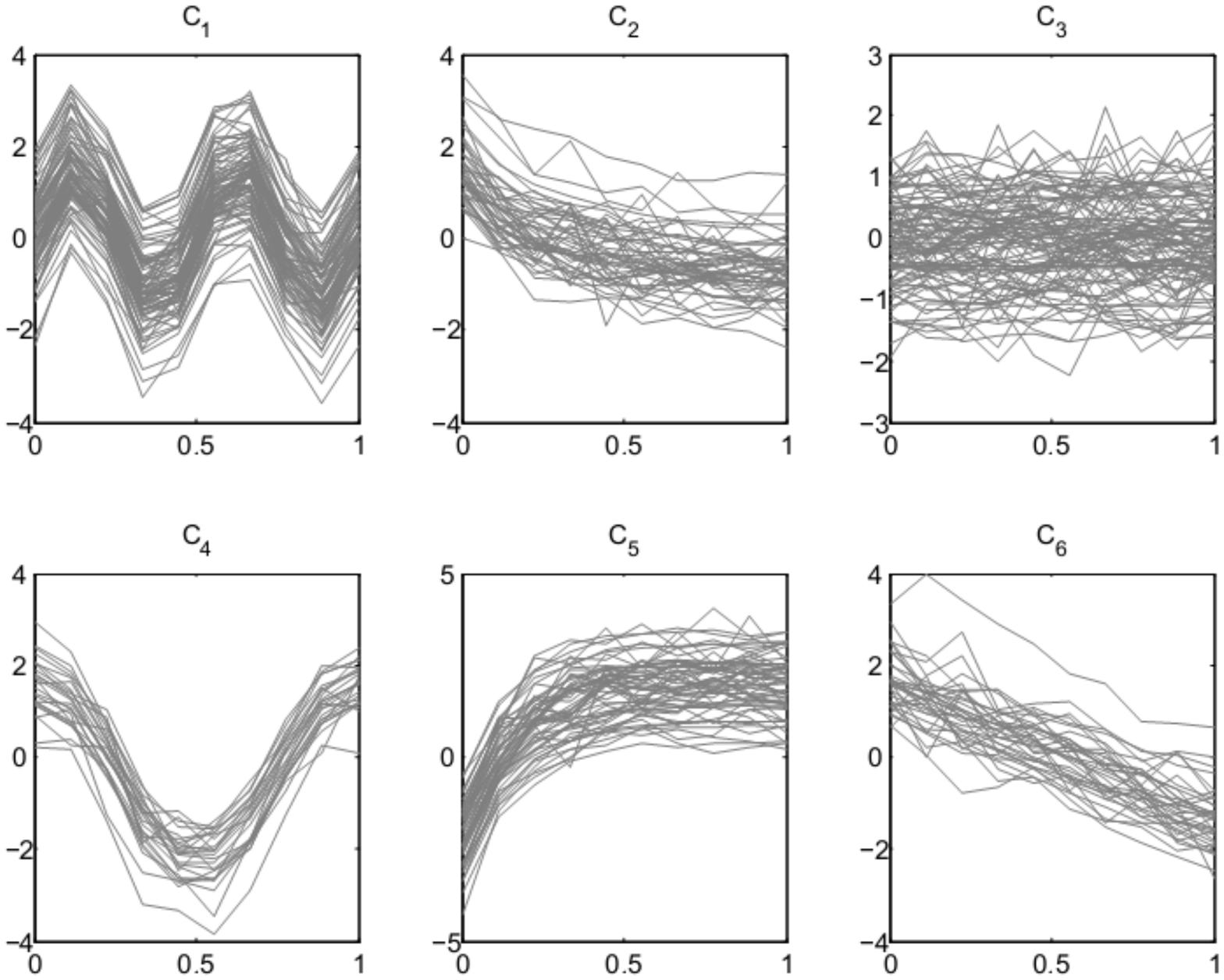}
\caption{Data set generated for simulation study.}
\label{fig.5.1 Simulated Dataset}
\end{figure}

As a first experiment, we generated $K=6$ clusters of numerical series at $n=10$ equally spaced time points in $[0,1]$ as described in Coffey \textit{et al.} (2014). Distinct cluster specific models were used (subscript $i$ refers to the series, subscript $j$ refers to the time domain):
\begin{eqnarray}
\nonumber && y_{ij}^{(1)}=\alpha_i +\sin(\beta_i*\pi *x_{ij}) + \gamma_{i} +\varepsilon_{ij} \\
\nonumber && y_{ij}^{(2)}=x_{ij} + (\delta_i)^{-3} + \iota_i + \gamma_{i} +\varepsilon_{ij}\\
\nonumber && y_{ij}^{(3)}=\nu_i + \gamma_{i}+\varepsilon_{ij}\\
\nonumber && y_{ij}^{(4)}=\zeta_i +\cos(\zeta_i*\pi *x_{ij}) + \gamma_{i} +\varepsilon_{ij}\\
\nonumber && y_{ij}^{(5)}=\xi_i - \eta_i * \exp (-\theta_i *x_{i}) + \gamma_{i} +\varepsilon_{ij}\\
\nonumber && y_{ij}^{(6)}= -3 (x_{ij}-0.5) + \gamma_{i} +\varepsilon_{ij}\\
\nonumber
\end{eqnarray}
where:\\
$\alpha_i \sim N(\sqrt{2};\sigma^2_e)$ with $\sigma^2_e=0.08$,
$\beta_i \sim N(4*\pi;\sigma^2_e)$,
$\delta_i \sim N(0.75;\sigma^2_e)$,\\
$\iota_i \sim N(1;\sigma^2_e)$,
$\nu_i \sim N(0;\sigma^2_e)$,
$\zeta_i \sim N(2;\sigma^2_e)$,
$\xi_i \sim N(2;\sigma^2_v)$ with $\sigma^2_v=0.85$,
$\eta_i \sim N(4;\sigma^2_v)$,
$\theta_i \sim N(6;\sigma^2_e)$,
$\gamma_{i}\sim N(0;\sigma^2_u)$ with $\sigma^2_u$ ranging from $0.3$ to $1$ and
$\varepsilon_{ij}$ is an autoregressive model of order 1.\\
Cluster means were chosen to reflect the situation where there are series that show little variation in value over time (as given by cluster 3) and series which have distinct signal over time.
Cluster sizes were equal to $90$, $50$, $100$, $25$, $60$ and  $35$, for cluster $1,2,3,4,5,6$ respectively, giving a total number of $360$ simulated series. Data set is plotted in Fig.\ref{fig.5.1 Simulated Dataset}.\\

\begin{center}
Figure \ref{fig.5.1 Simulated Dataset} about here.\\
\end{center}

\noindent Given the nature of the simulated series, we are interested in the similarity of the shape of the series. For this reason the chosen metric was the Penrose shape distance (Penrose, 1952), defined as:

\begin{eqnarray}
\label{PSD}
d_{i,j}=\sqrt{\frac{n_i}{n_i-1} (d_{i,j}^2-q_{ij}^2) },
\end{eqnarray}
where $d_{i,j}^2$ is the squared average Euclidean distance coefficient and $q_{ij}^2=\frac{1}{n_i^2}  \left(\sum_{j=i}^{n_i}y_{ji}-\sum_{j=1}^{n_i}c_{jk}\right)^2$.

\begin{figure}[h]
\centering
\includegraphics[width=1.00\textwidth]{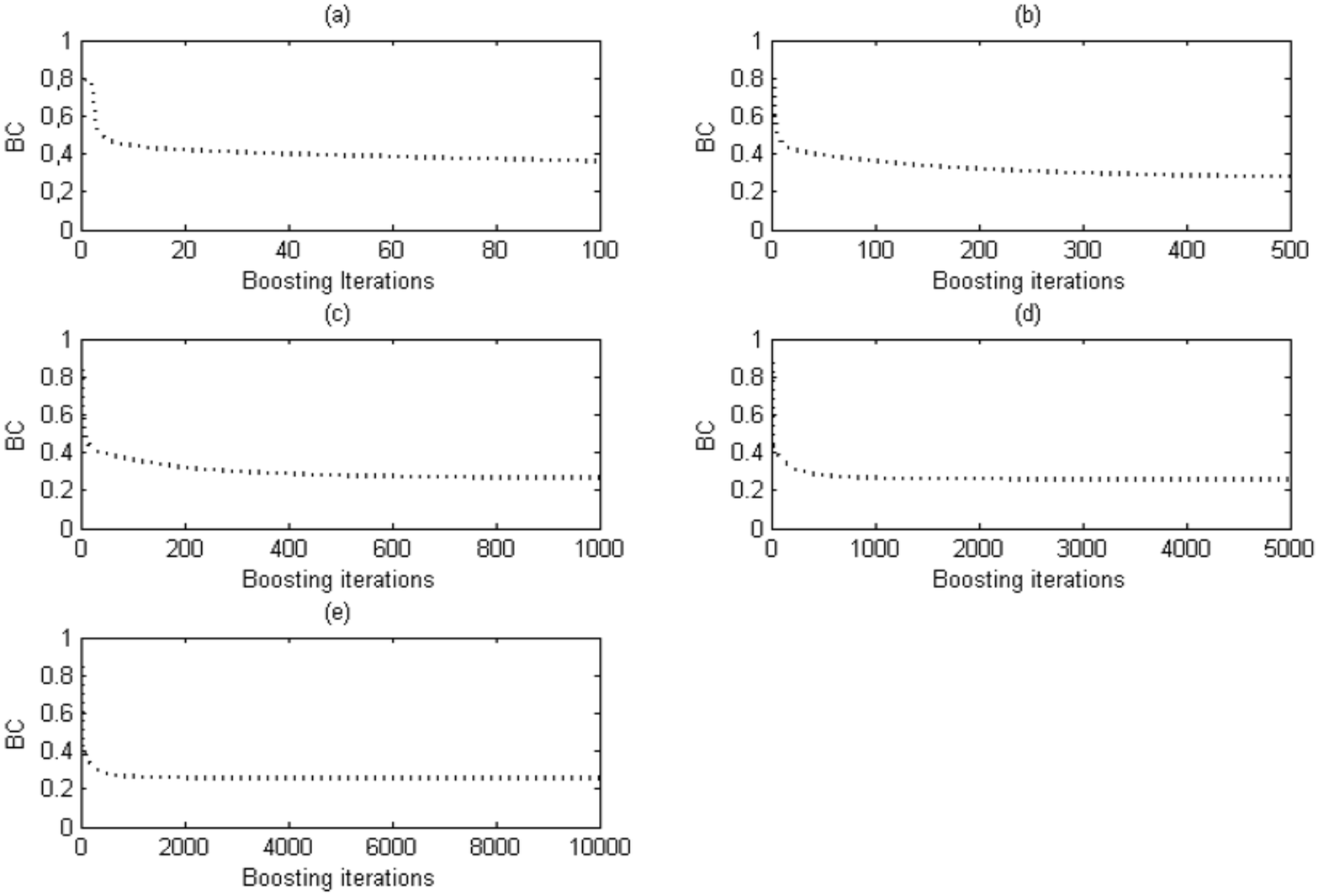}
\caption{$BC$ function progress through $100$ boosting iterations: (a) = $100$ boosting iterations; (b) = $500$ boosting iterations; (c) = $1000$ boosting iterations; (d) = $5000$ boosting iterations; (e) = $10000$ boosting iterations.}
\label{fig.5.2 IT1}
\end{figure}

We performed five analyses with $100, 500, 1000, 5000$ and $10000$ boosting iterations. In all cases we set $10$ random starting points. Figure \ref{fig.5.2 IT1} shows the behavior of the BC function as defined in equation \eqref{solpd} during the boosting iterations. In this case the $BC$ values appear to be non-increasing as the number of iterations increases. The values of the $BC$ function are equal to  $0.3615, 0.2783, 0.2643, 0.2584, 0.2583$ for $100, 500, 1000, 5000$ and $10000$ boosting iterations respectively.\\

\begin{center}
Figure \ref{fig.5.2 IT1} about here.\\
\end{center}

\begin{figure}[h]
\centering
\includegraphics[width=1.00\textwidth]{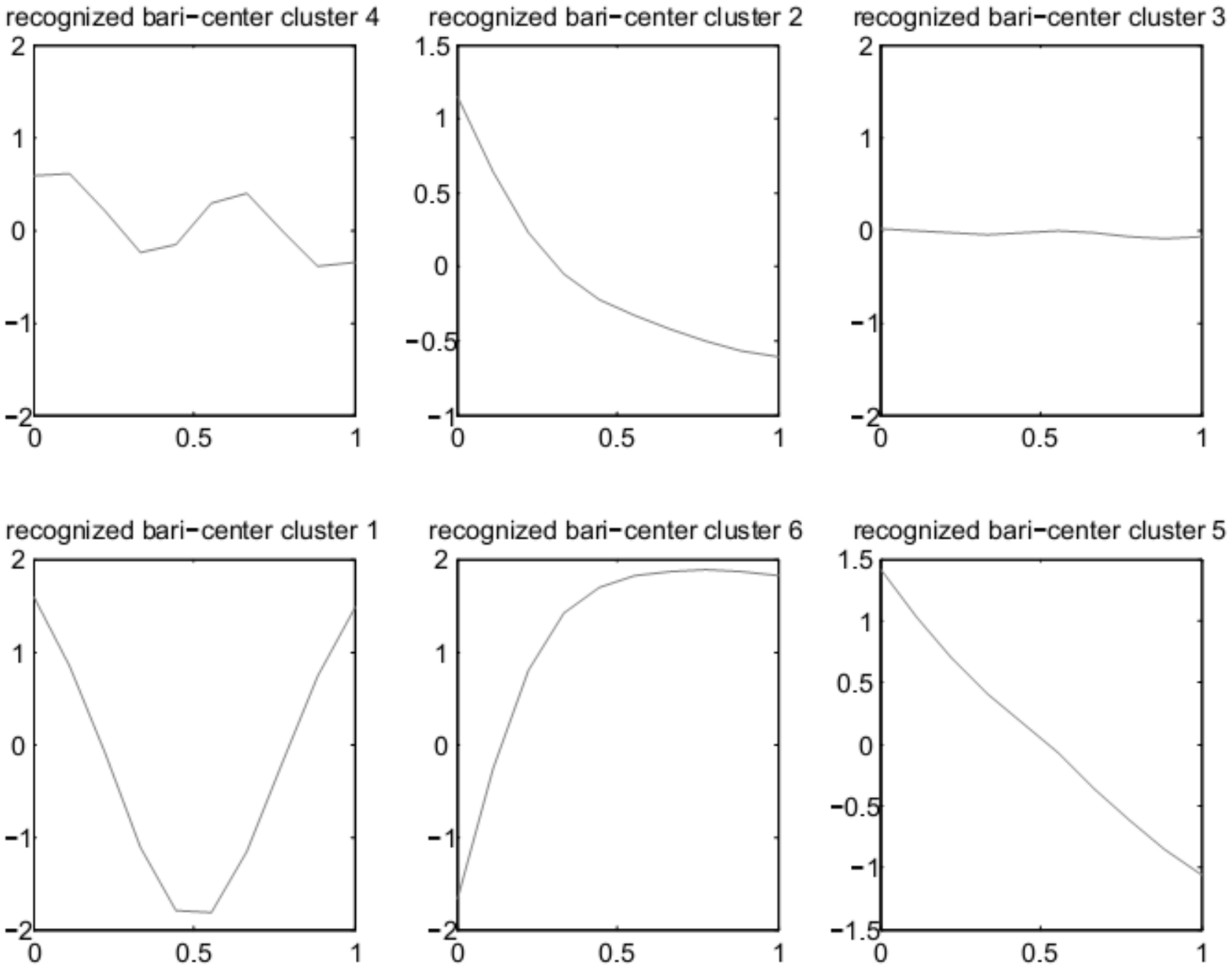}
\caption{Estimated cluster centers.}
\label{fig.5.3 ICR}
\end{figure}

\noindent All the solutions return in fact the same results in terms of estimated centers: in example, figure \ref{fig.5.3 ICR} shows the estimated cluster centers for each cluster as returned by the first analysis. \\

\begin{center}
Figure \ref{fig.5.3 ICR} about here.\\
\end{center}

\noindent For this data set, by using the Penrose shape distance, the Fuzzy Rand Index is equal to $0.8599, 0.8954, 0.9059, 0.9178$ and $0.9194$ for the solutions with respectively $100, 500, 1000, 5000$ and $10000$ boosting iterations. Even if the solutions in terms of "hard" clustering are the same, the difference in terms of fuzzy rand index indicates that the partitions returned by the proposed algorithm are really close to the true one.  The true value of the $BC$ index is $0.1977$. 

\subsection{Synthetic data set}

\begin{figure}[h]
\centering
\includegraphics[width=1.00\textwidth]{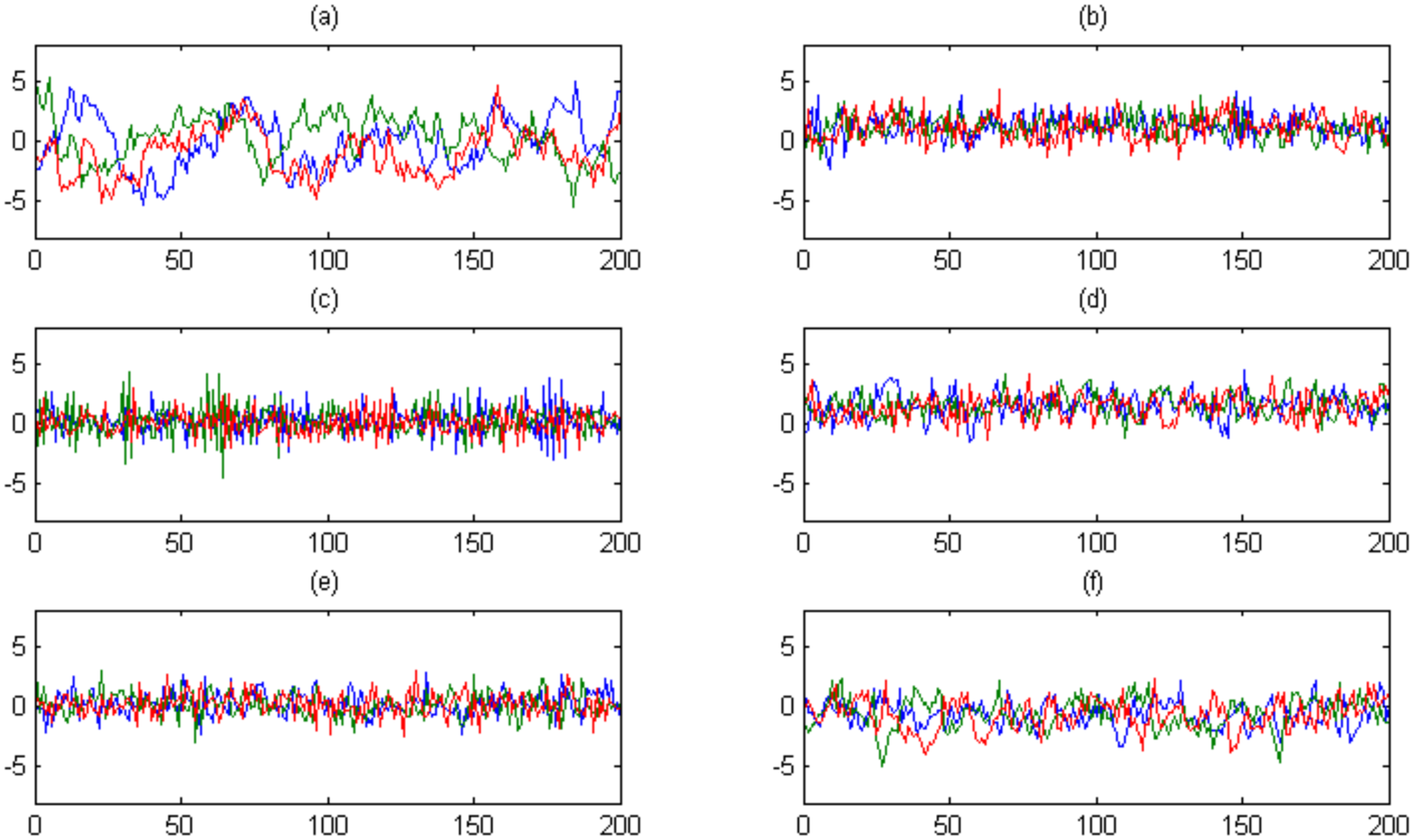}
\caption{\footnotesize Synthetic.tseries data set.}
\label{syntdata}
\end{figure}

Synthetic.tseries data set is freely available from the \texttt{TSclust} R-package (Montero and Vilar, 2014).
Synthetic.tseries data consist of three partial realizations of length $n = 200 $ of six first order autoregressive models. Figure \ref{syntdata} shows separately the six groups of series. \\

\begin{center}
Figure \ref{syntdata} about here.
\end{center}

\noindent Subplot (a) shows an AR(1) process with moderate autocorrelation. Subplot (b) contains series from a bi-linear process with approximately quadratic conditional mean. Subplot (c) is formed by an exponential autoregressive model with a more complex non-linear structure. Subplot (d) shows a self-exciting threshold autoregressive model with a relatively strong non-linearity. Subplot (e) contains series generated by a general non-linear autoregressive model and subplot (f) shows a smooth transition autoregressive model presenting a weak non-linear structure. As we did not generated these series we do not show completely the simulation setting. For more details about the generating models we refer to Montero and Vilar (2014), pag. 24.\\
Assuming that the aim of cluster analysis is to discover the similarity between underlying models, the "true" cluster solution is given by the six clusters involving the three series from the same generating model.
Given the nature of the data set considered, we use a periodogram-based distance measure proposed by Caiado \textit{at al.} (2006). It assesses the dissimilarity between the corresponding spectral representation of time series.

\noindent By following also the suggestion of to Montero and Vilar (2014), an interesting alternative to measure the dissimilarity between time series is the frequency domain approach. Power spectrum analysis is concerned with the distribution of the signal power in the frequency domain. The power-spectral density is defined as the Fourier transform of the autocorrelation function of $i-th$ series. It is a measure of self-similarity of a signal with its delayed version. The classic method for estimation of the power spectral density of an $n$-sample record is the periodogram introduced by Schuster (1897). 
Let $y$ and $y^{'}$ be two time series of length $n$.\\
Let $f_j=2\pi j/n$, $ j=1,\ldots, n/2$ in the range $0$ to $\pi$, be the frequencies of the series.\\
Let ${{PSD}}_{y}(f_j)=\frac{1}{n}\sum_{t=1}^{n}| y_t(f_j) \exp{(-\iota t f_j)}|^2$ and ${{PSD}}_{y^{'}}(f_j)=\frac{1}{n}\sum_{t=1}^{n}| y^{'}_t(f_j) \exp{(-\iota t f_j)}|^2$ be the periodograms of series $y$ and $y^{'}$, respectively.\\
\noindent Finally, the dissimilarity measure between  $y$ and $y^{'}$ proposed by Caiado \textit{et al.} (2006) is defined as the Euclidean distance between periodogram ordinates :
\begin{eqnarray}\label{EuclPSD}
d_{y,y^{'}}=\sqrt{\sum_{j=1}^{(n/2)} [{{PSD}}_y(f_j)-{{PSD}}_{y^{'}}(f_j)]^2}.
\end{eqnarray}
We performed our analysis by setting $800$ boosting iterations and $10$ random starting points.\\
\noindent Table \ref{Result Synthetic} shows the results of applying our algorithm to the Synthetic.tseries data set. Each series is assigned to the estimated cluster according to the value of the membership probability matrix. In order to obtain the Fuzzy Rand Index, we computed the true cluster centers with a periodogram modeled by P-spline , and then we computed the true probabilities by applying equation \eqref{probs} by using the periodogram-based distance as in equation \eqref{EuclPSD}.\\
The Fuzzy Rand is equal to $ 0.9698 $. Even if the solutions in terms of "hard" clustering seems to be excellent (since only series is misclassified), the difference in terms of Fuzzy Rand index indicates that the partitions returned by the algorithm are really close to the true one. \\

\begin{center}
Table \ref{Result Synthetic} about here.
\end{center}

\subsection{A real data example}
\label{sec5.4}

\begin{figure}[h]
\centering
\includegraphics[width=1.00\textwidth]{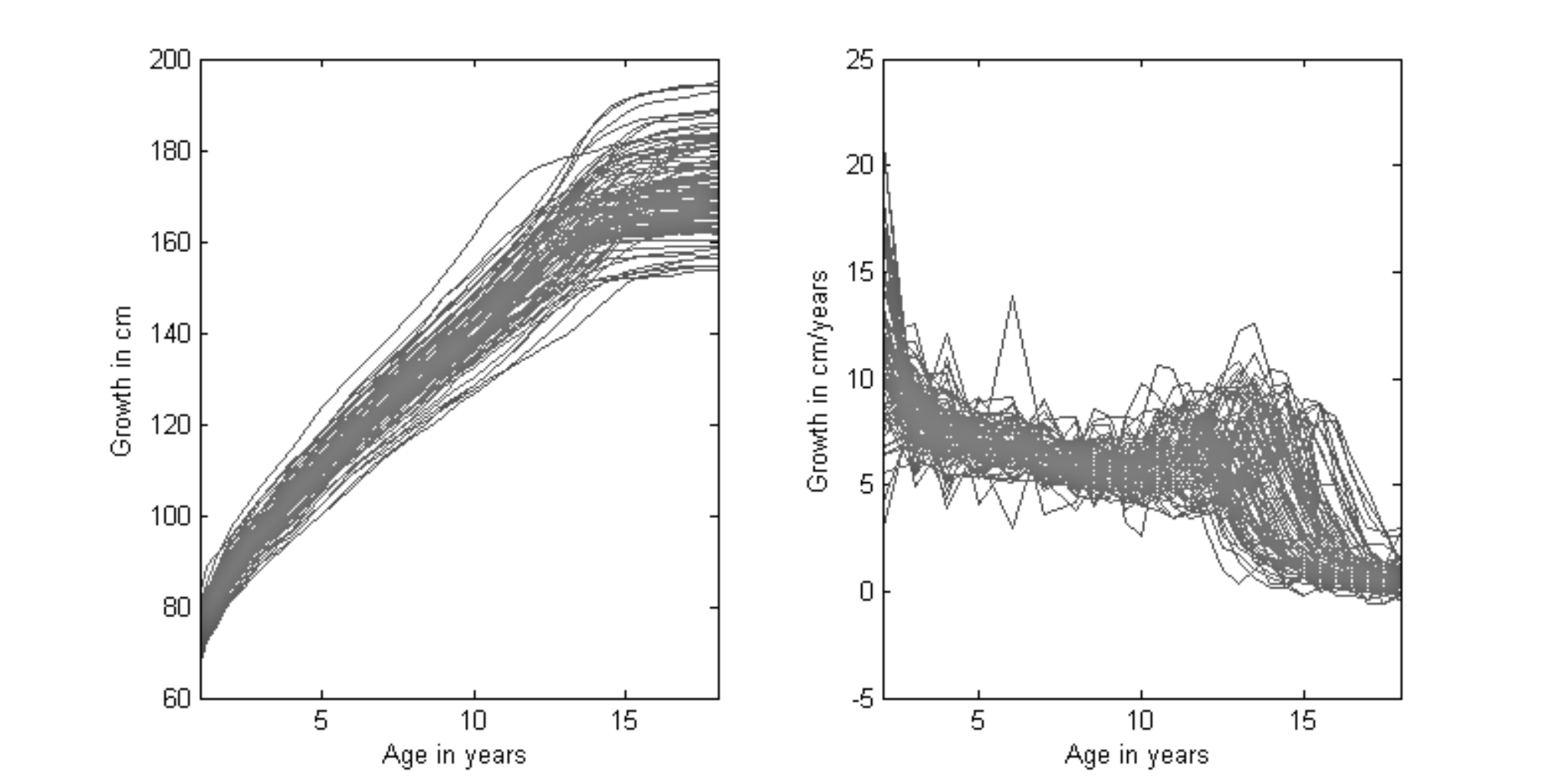}
\caption{Growth curves (left hand side) and growth velocity curves (right hand side) of 93 children from Berkeley Growth Study data.}
\label{fig.growth}
\end{figure}

The "growth" data set is freely available from the internal repository of the R-package \textit{fda} (Ramsay \textit{et al.}, 2012). This data set comes from the Berkeley Growth Study (Tuddenham and Snyder, 1954). Left hand side of figure \ref{fig.growth} shows the growth curves of 93 children, 39 boys and 54 girls, starting by the age of one year till the age of 18. The right hand side of the same figure displays the corresponding growth velocities.  \\

\begin{center}
Figure \ref{fig.growth} about here.
\end{center}

\begin{figure}[h]
\centering
\includegraphics[width=1.00\textwidth]{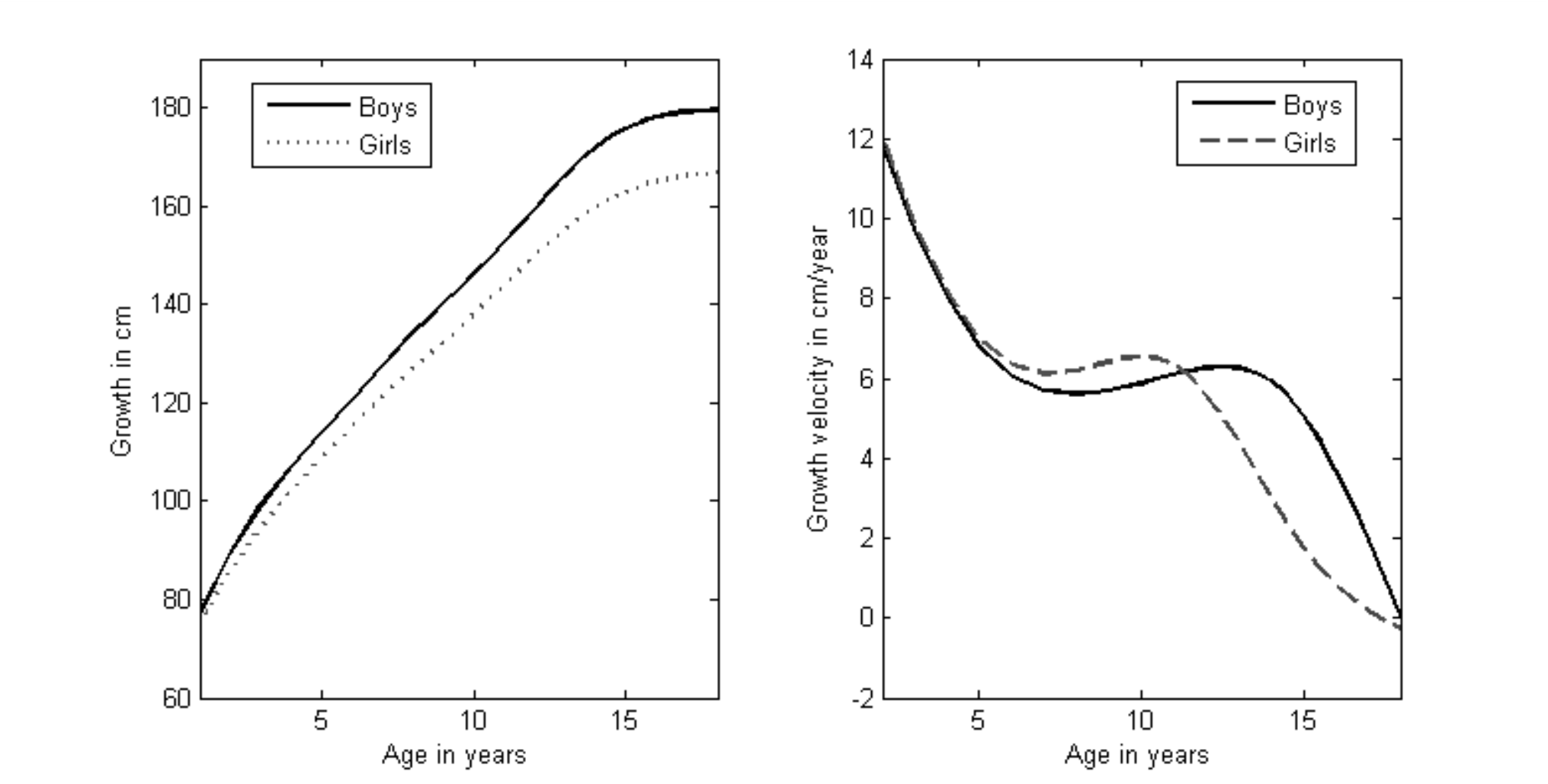}
\caption{Estimated centers of growth curves (left hand side) and growth velocities (right hand side): Euclidean distance.}
\label{fig.growth_Eucl}
\end{figure}

\noindent In the framework of cluster analysis this data set was mainly used for problems of clustering of misaligned data (Sangalli \textit{et al.}, 2010a, 2010b). We performed two analyses with 800 boosting iterations and with 10 random starting point with $k = 2$. In the first partitioning analysis we used the Euclidean distance. The estimated centers of both the growth curves and the growth velocity curves are displayed respectively in the left and right hand side of figure \ref{fig.growth_Eucl}. As it can be noted, Euclidean distance discriminates between children growing more and children growing less. This can be appreciated by looking at left hand side of the same figure. On average, as expected, boys grow more than girls. \\

\begin{center}
Figure \ref{fig.growth_Eucl} about here.
\end{center}

\begin{figure}[h]
\centering
\includegraphics[width=1.00\textwidth]{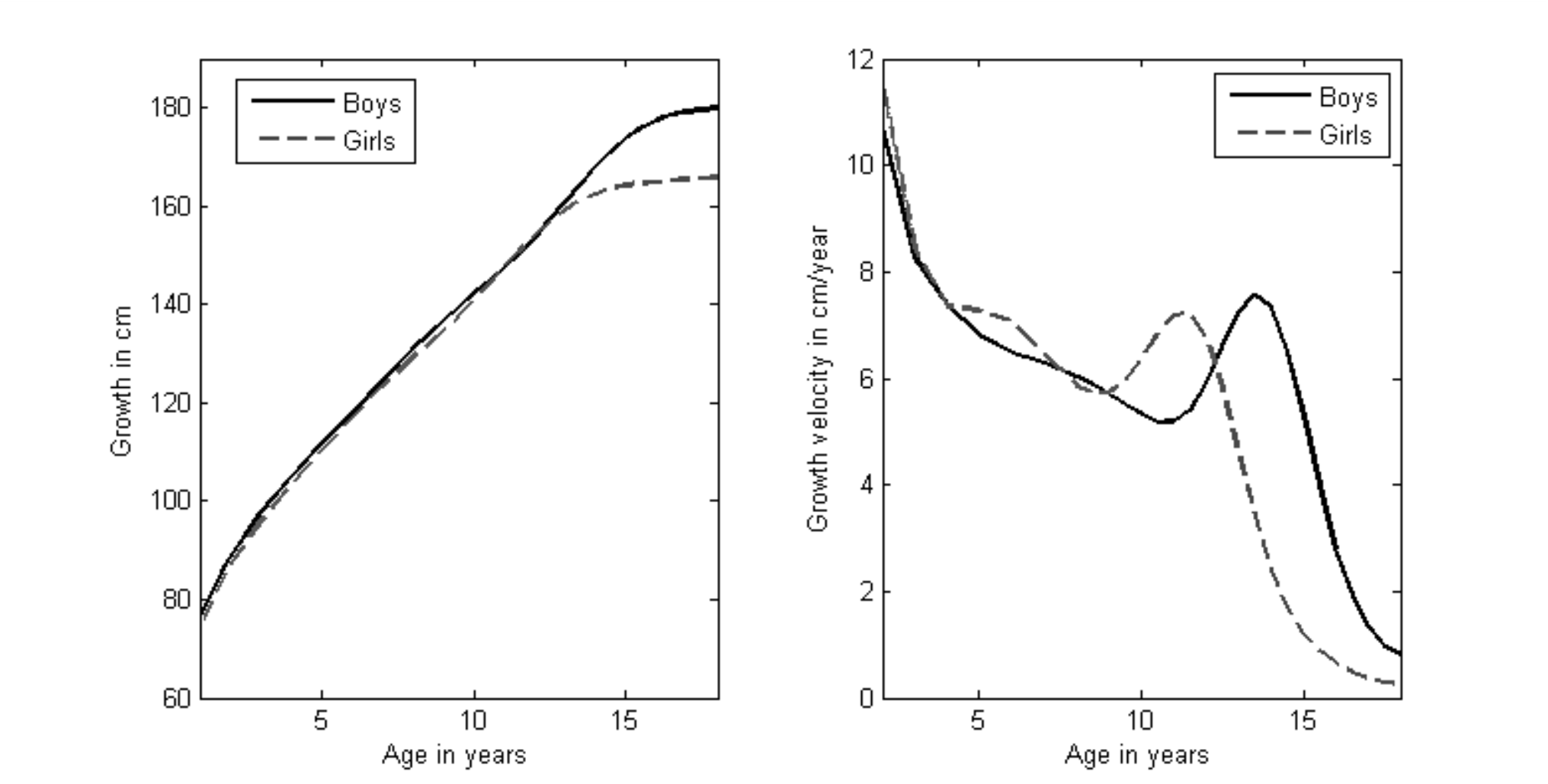}
\caption{Estimated centers of growth curves (left hand side) and growth velocities (right hand side): Penrose shape distance.}
\label{fig.growth_SD}
\end{figure}

\noindent Nevertheless, Euclidean distance does not seem the right measure to be used in such a case. Probably researchers are interested in the shape of both growth and growth velocity curves during the years. For this reason, we repeated the analysis by using the Penrose shape distance as defined in equation \eqref{PSD}. Figure \ref{fig.growth_SD} shows the estimated centers for both the growth and the growth velocity curves. The recognized centers are really similar to the ones obtained by Sangalli \textit{et al.} (2010a; 2010b): firstly, as confirmed by looking at tables \ref{CF3} and \ref{CF4} with respect to tables \ref{CF1} and \ref{CF2}, there is a neat separation of boys and girls. Secondly, by looking at right hand side of figure \ref{fig.growth_SD}, boys start to grow later but they seem to have a more pronounced growth, as it can be noticed by looking at the higher peak in correspondence of 15 year. \\

\begin{center}
Figure \ref{fig.growth_SD} about here.
\end{center}

\begin{center}
Table \ref{CF1} about here.
\end{center}

\begin{center}
Table \ref{CF2} about here.
\end{center}

\begin{center}
Table \ref{CF3} about here.
\end{center}

\begin{center}
Table \ref{CF4} about here.
\end{center}

\noindent The Fuzzy Rand index is equal to $0.8884$ and $0.8240$ by using the Euclidean distance for the partitions of growth and growth velocity curves respectively. The Fuzzy Rand index is equal to $1.000$ and $0.9246$ by using the Penrose shape distance for the partitions of growth and growth velocity curves respectively.

\section{Concluding remarks}
\label{sec5.4} 
In this paper we merged two approaches, theoretically motivated for respectively unsupervised and supervised classification cases, to propose a new non-hierachical fuzzy clustering algorithm. \\
\indent From the Probabilistic Distance (PD) clustering (Ben-Israel and Iyigun, 2008) approach we shared the idea of determining the probabilities of each series to any of the $k$ clusters. As this probability is directly related to the distance of each series from the cluster centers, there are no degrees of freedom in determine the membership matrix. \\
\indent From the Boosting approach (Freund and Schapire, 1997) we shared the idea of weighting each series according some measure of badness of fit in order to define an unsupervised learning process based on a weighted resampling procedure. In contrast to the boosting approach, the higher the probability of a given instance to be member of a given cluster, the higher the weight of that instance in the resampling process. As a learner we can use any smoothing spline technique. We used a P-spline smoother (Eilers and Marx, 1996) because of its nice properties and we choose the optimal spline parameter with the V-curve criterion as defined by Frasso and Eilers (2015).  In this way we defined a suitable loss function and, at the same time, we proposed a fuzzy clustering procedure that does not depend on the definition of a fuzzifier parameter. \\
\indent To evaluate the performance of our proposal, we conducted three experiments, one of them on simulated data and the remaining two on data sets known in literature. The results show that our Boosted-oriented procedure show good performance in terms of data partitioning. Even if the final fuzzy partition is sensitive to the choice of a distance measure, it is independent on any other input parameters. This consideration allows to define a suitable true fuzzy partition with which evaluate the final solution in terms of Fuzzy Rand Index (H\"ullermeier \textit{et al.}, 2012). The weigthed re-sampling process allows each series to contribute to the composition of each cluster as well as the adaptive estimation of cluster centers allows the algorithm to learn by its progresses. \\
It is worth-nothing that, as in any partitioning problem, the choice of the distance measure can influence the goodness of partition.

\vspace{\fill}\pagebreak

\section*{Figures}
\vskip1pt

\vspace{\fill}\pagebreak

\vspace{\fill}\pagebreak

\vspace{\fill}\pagebreak

\vspace{\fill}\pagebreak

\vspace{\fill}\pagebreak

\vspace{\fill}\pagebreak

\vspace{\fill}\pagebreak

\section*{Tables}
\vskip1pt

\begin{table}[h!]
\footnotesize
    \centering
		\caption{\footnotesize Confusion matrix from clustering on Synthetic.tseries data set.}
\begin{tabular}{cc|c|c|c|c|c|c|c}
\cline{3-8}
& & \multicolumn{6}{ c| }{Estimated Clusters} \\ \cline{3-8}
& & C1 & C2 & C3 & C4 & C5 & C6  \\ \cline{1-8}
\multicolumn{1}{ |c  }{\multirow{6}{*}{True Clusters} } &
\multicolumn{1}{ |c| }{a} & 0 & 0 & 0 & 0 & 0 & 3 &  \\ \cline{2-8}
\multicolumn{1}{ |c  }{}      &
\multicolumn{1}{ |c| }{b} & 0 & 1 & 0 & 2 & 0 & 0 & \\ \cline{2-8}
\multicolumn{1}{ |c  }{} &
\multicolumn{1}{ |c| }{c} & 3 & 0 & 0 & 0 & 0 & 0 &     \\ \cline{2-8}
\multicolumn{1}{ |c  }{} &
\multicolumn{1}{ |c| }{d} & 0 & 3 & 0 & 0 & 0 & 0 &    \\ \cline{2-8}
\multicolumn{1}{ |c  }{} &
\multicolumn{1}{ |c| }{e} & 0 & 0 & 3 & 0 & 0 & 0 &     \\ \cline{2-8}
\multicolumn{1}{ |c  }{} &
\multicolumn{1}{ |c| }{f} & 0 & 0 & 0 & 0 & 3 & 0 &    \\ \cline{1-8}
\end{tabular}
    \label{Result Synthetic}
\end{table}


\begin{table}[h!]
\centering
\caption{Confusion matrix of growth curves with the Euclidean distance. Series have been assigned to the clusters according the values of membership probabilities computed as in equation \eqref{probs}. }
\begin{tabular}{l|l|l|l|}
\cline{2-4}
 &  & Cluster 1 & Cluster 2 \\ 
\cline{2-4}
 & Boys & \multicolumn{1}{r|}{23} & \multicolumn{1}{r|}{16} \\ 
\cline{2-4}
 & Girls & \multicolumn{1}{r|}{16} & \multicolumn{1}{r|}{38} \\ 
\cline{2-4}
\end{tabular}
\label{CF1}
\end{table}


\begin{table}[h!]
\centering
\caption{Confusion matrix of growth velocity curves with the Euclidean distance. Series have been assigned to the clusters according the values of membership probabilities computed as in equation \eqref{probs}. }
\begin{tabular}{l|l|l|l|}
\cline{2-4}
 &  & Cluster 1 & Cluster 2 \\ 
\cline{2-4}
 & Boys & \multicolumn{1}{r|}{31} & \multicolumn{1}{r|}{8} \\ 
\cline{2-4}
 & Girls & \multicolumn{1}{r|}{9} & \multicolumn{1}{r|}{45} \\ 
\cline{2-4} 
\end{tabular}
\label{CF2}
\end{table}


\begin{table}[h!]
\centering
\caption{Confusion matrix of growth curves with the Penrose shape distance. Series have been assigned to the clusters according the values of membership probabilities computed as in equation \eqref{probs}. }
\begin{tabular}{l|l|l|l|}
\cline{2-4}
 &  & Cluster 1 & Cluster 2 \\ 
\cline{2-4}
 & Boys & \multicolumn{1}{r|}{0} & \multicolumn{1}{r|}{39} \\ 
\cline{2-4}
 & Girls & \multicolumn{1}{r|}{52} & \multicolumn{1}{r|}{2} \\ 
\cline{2-4}
\end{tabular}
\label{CF3}
\end{table}


\begin{table}[h!]
\centering
\caption{Confusion matrix of growth velocity curves with the Penrose shape distance. Series have been assigned to the clusters according the values of membership probabilities computed as in equation \eqref{probs}. }
\begin{tabular}{l|l|l|l|}
\cline{2-4}
 &  & Cluster 1 & Cluster 2 \\ 
\cline{2-4}
 & Boys & \multicolumn{1}{r|}{36} & \multicolumn{1}{r|}{3} \\ 
\cline{2-4}
 & Girls & \multicolumn{1}{r|}{4} & \multicolumn{1}{r|}{49} \\ 
\cline{2-4}
\end{tabular}
\label{CF4}
\end{table}

\end{document}